\documentclass[a4paper]{jpconf}
\usepackage{graphicx}
\usepackage{amsmath}
\usepackage[amssymb, mediumspace]{SIunits}
\usepackage{multicol}
\usepackage[
  font={small,sl},
  labelfont={sc,bf},
  labelsep={space},
  format={plain},
  justification={justified},
  singlelinecheck={yes},
  figureposition={bottom},
  tableposition={top},
  margin={4em}
  ]{caption}[2008/04/01]
\usepackage{color}
  \definecolor{my_color_for_extrainfo}{rgb}{0.4,0.2,0.2}
  \definecolor{myred}{rgb}{0.9,0.1,0.1}
  \definecolor{myDarkRed}{rgb}{0.5,0,0}       % used for CITATIONS (see hyperref package).
  \definecolor{myDarkBlue}{rgb}{0,0,0.6}      % used for LINKS and URLs (see hyperref package).
  \definecolor{myDarkGreen}{rgb}{0,0.4,0}     % used for REFERENCES (see hyperref package).
  \definecolor{myDarkGray}{rgb}{0.4,0.4,0.4}  % used for page margin notes
  \definecolor{black}{rgb}{0,0,0}
  \definecolor{white}{rgb}{1,1,1}
\usepackage{ifthen}
  \newcommand{\showextrainfo}{1}
  \newcommand{\extrainfo}[1]{%
    {\color{my_color_for_extrainfo}%
      \ifthenelse{\isundefined{\showextrainfo}}{}{#1}%
    }%
  }

  \newcommand{\inred}[1]{{#1}}

\usepackage[
  a4paper={true},
  colorlinks={true},
  linkcolor={myDarkBlue},
  citecolor={myDarkGreen},
  urlcolor={myDarkBlue},
  hyperfootnotes={false}
  ]{hyperref}

\begin{document}

\title{Non-linear Simulations of MHD Instabilities in Tokamaks
Including Eddy Current Effects and Perspectives for the
Extension to Halo Currents}

\author{M.~Hoelzl$^1$, G.T.A.~Huijsmans$^2$, P.~Merkel$^1$, C. Atanasiu$^3$, K. Lackner$^1$, E. Nardon$^4$, K. Aleynikova$^2$, F. Liu$^2$, E. Strumberger$^1$, R. McAdams$^{5,6}$, I. Chapman$^6$, A. Fil$^4$}

\address{
$^1$ Max-Planck-Institute for Plasmaphysics, Boltzmannstr. 2, 85748 Garching, Germany \\
$^2$ ITER Organisation, Route de Vinon sur Verdon, St-Paul-lez-Durance, France \\
$^3$ Association EURATOM-MEdC Romania, P.O. Box MG-36, Magurele, Bucharest, Romania \\
$^4$ CEA, IRFM, CEA Cadarache, F-13108 St Paul-lez-Durance, France \\
$^5$ York Plasma Institute, University of York, York, YO10 5DD, UK \\
$^6$ Culham Science Centre, Abingdon, OX14 3DB, UK
}

\ead{mhoelzl@ipp.mpg.de}

\inred{
\begin{abstract}
The dynamics of large scale plasma instabilities can be strongly influenced
by the mutual interaction with currents flowing in conducting vessel structures.
Especially eddy currents caused by time-varying magnetic perturbations and
halo currents flowing directly from the plasma into the walls are important.

The relevance of a resistive wall model is directly evident for Resistive
Wall Modes (RWMs) or Vertical Displacement Events (VDEs). However, also the linear
and non-linear properties of most other large-scale instabilities may be
influenced significantly by the interaction with currents in conducting
structures near the plasma. The understanding of halo currents arising
during disruptions and VDEs, which are a serious concern for ITER as they
may lead to strong asymmetric forces on vessel structures, could also benefit
strongly from these non-linear modeling capabilities.

Modeling the plasma dynamics and its interaction with wall currents requires
solving the magneto-hydrodynamic (MHD) equations in realistic toroidal X-point
geometry consistently coupled with a model for the vacuum region and the
resistive conducting structures. With this in mind, the non-linear finite
element MHD code JOREK~\cite{Huysmans2007,Czarny2008} has been coupled~\cite{Hoelzl2012A}
with the resistive wall code STARWALL~\cite{Merkel2006}, which allows us to include the
effects of eddy currents in 3D conducting structures in non-linear MHD
simulations.

This article summarizes the capabilities of the coupled JOREK-STARWALL system and
presents benchmark results as well as first applications to non-linear simulations of RWMs,
VDEs, disruptions triggered by massive gas injection, and Quiescent H-Mode. As an
outlook, the perspectives for extending the model to halo currents are described.
\end{abstract}
}

% ==============================================================================
\section{Introduction}
% ==============================================================================

Large scale plasma instabilities can be strongly influenced by the interaction with
currents in conducting structures. The resulting forces on wall structures are an
important topic for ITER as well.
In order to address those questions involving non-linear plasma
dynamics and its interaction with wall currents requires a code that solves the
magneto-hydrodynamic (MHD) equations in realistic toroidal X-point
geometry consistently coupled with a model for the vacuum region and the
resistive conducting structures. With this in mind, the non-linear finite
element MHD code JOREK~\cite{Huysmans2007,Czarny2008} has been coupled~\cite{Hoelzl2012A}
with the resistive wall code STARWALL~\cite{Merkel2006}, which allows us to include the
effects of eddy currents in 3D conducting structures in non-linear MHD
simulations.

This article summarizes the capabilities of the coupled JOREK-STARWALL system (Section~\ref{:jorstar})
and presents benchmark results (Section~\ref{:bench}) as well as first physics applications to non-linear
simulations (Section~\ref{:phys}) of resistive wall modes (RWMs), vertical displacement events (VDEs),
disruptions triggered by massive gas injection (MGI), and quiescent
H-Mode (QH mode). This publication only provides a brief overview for the physics investigations
and refers to separate papers going into more detail on each subject.
As an outlook, the perspectives for extending the model
to halo currents are described (Section~\ref{:halo}).

% ==============================================================================
\section{The Model}\label{:jorstar}
% ==============================================================================

% ==============================================================================
\subsection{The non-linear MHD code JOREK}\label{:jorstar:jorek}
% ==============================================================================

The non-linear MHD code JOREK~\cite{Huysmans2007} solves a set of reduced MHD
equations~\cite{Strauss1997,Franck2015} in realistic toroidal X-point geometry. Two-fluid
extensions (diamagnetic drift, separate electron and ion temperatures) and a model
for deuterium neutrals exist. A full MHD model is available~\cite{Haverkort2013}, but
needs to be extended by X-point boundary conditions and two-fluid terms.

% ==============================================================================
\subsection{The resistive wall code STARWALL}\label{:jorstar:starw}
% ==============================================================================

The standalone STARWALL code is a resistive wall code~\cite{Merkel2006,Merkel2015}
capable of linear stability analysis of MHD modes in the presence of 3D resistive
walls. Since kinetic energy is neglected, its validity is restricted to cases where
plasma inertia does not play an important role. Resistive walls are discretized by triangles using the thin-wall
approximation. The validity of this approximation is discussed for instance in
Reference~\cite{Gimblett1986} and~\ref{:app:thin}.

Parts of STARWALL are currently implemented into the linear MHD code
CASTOR~\cite{Huysmans1993,Kerner1998,Strumberger2011A} along with additional
extensions to create the new code CASTOR3D~\cite{Strumberger2014} for linear studies
including coupling of different toroidal modes by 3D resistive walls, kinetic energy, plasma resistivity and plasma
rotation. A modified version of STARWALL (sometimes called STARWALL\_J) has been
coupled to JOREK to allow non-linear simulations with 3D resistive wall effects. This
coupling is described in the following Section.

\begin{figure}
\centering
  \includegraphics[height=0.3\textwidth]{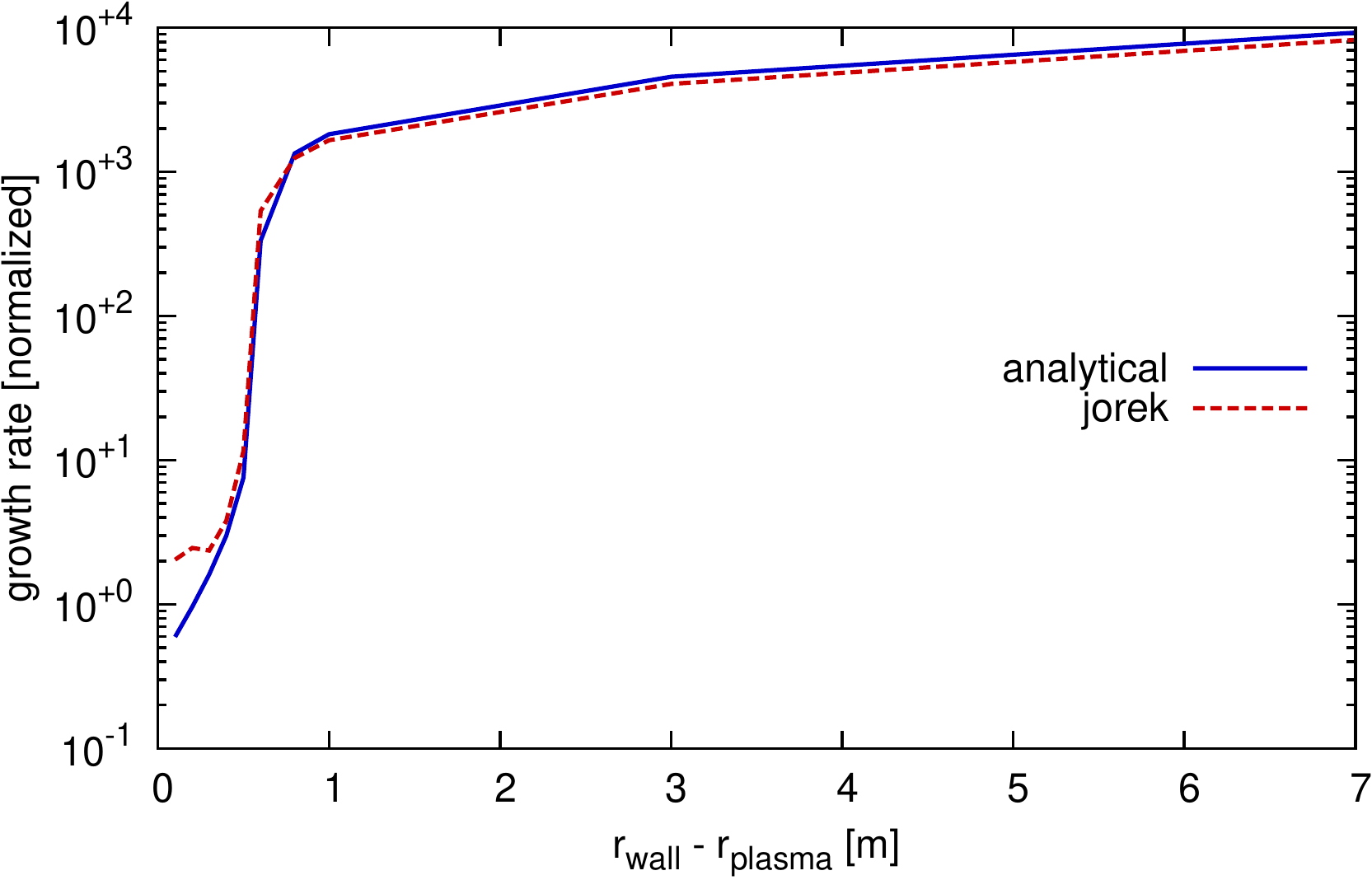}
\caption{Comparison of linear growth rates between analytical theory and JOREK-STARWALL results
for an RWM case in a circular limiter plasma with minor radius $a=1\;\mathrm{m}$, major radius
$R=10\;\mathrm{m}$ and at
a wall resistivity of $\eta_\text{wall}=2.5\cdot10^{-5}\;\Omega\;\mathrm{m}$. }
\label{fig:rwm-bench}
\end{figure}

% ==============================================================================
\subsection{The JOREK-STARWALL Coupling}\label{:jorstar:coupling}
% ==============================================================================

The fully implicit JOREK-STARWALL coupling via a natural boundary condition has
already been described previously in detail~\cite{Hoelzl2012A}. This Section only
contains a brief summary.

The boundary of the JOREK computational domain is used as ``interface'' for the
coupling of JOREK and STARWALL. STARWALL solves the field equations outside
the interface (vacuum region and resistive wall structures) by a variational
ansatz with the poloidal magnetic flux $\psi$ as boundary condition which allows
us to obtain an expression for the time evolution of wall currents. STARWALL computes
a set of matrices representing this time evolution equation, which are used
inside JOREK to keep track of the wall currents.

By considering all possible ``unit perturbations'', STARWALL also calculates
matrices which allow to express the tangential magnetic field at the JOREK
boundary in terms of wall currents and poloidal magnetic flux at the interface.
This correlation is used inside JOREK to compute the natural boundary condition
-- a surface integral in the current definition equation originating from partial
integration.

The implementation in JOREK is performed such that the full implicitness
of the code is preserved without introducing additional unknowns. To achieve
this, the unknown wall currents at time step $n+1$ are expressed in terms of
the (as well unknown) JOREK variables at the same time step. The coupling is
currently implemented only for the reduced MHD model, where the tangential
magnetic field perturbation has no toroidal component. STARWALL and the
coupling need to be extended in the future by additional matrices to add this
additional component. As the STARWALL representation of wall currents allows
only divergence-free surface currents, halo currents can presently not be
included. Consequently, simulations are limited to eddy currents.

JOREK-STARWALL has already been benchmarked successfully against analytical
theory and linear codes as briefly shown in Section~\ref{:bench}. First physics
results obtained with JOREK-STARWALL are presented in Section~\ref{:phys}.
An outlook to the inclusion of halo currents, which will be especially important
for disruption simulations, is given in Section~\ref{:halo}. In separate projects,
extensions to impurity massive gas injection and runaway electrons are currently
started such that we will be able to address the most important questions related
to disruption physics in the future.

\begin{figure}
\centering
\includegraphics[width=0.25\textwidth]{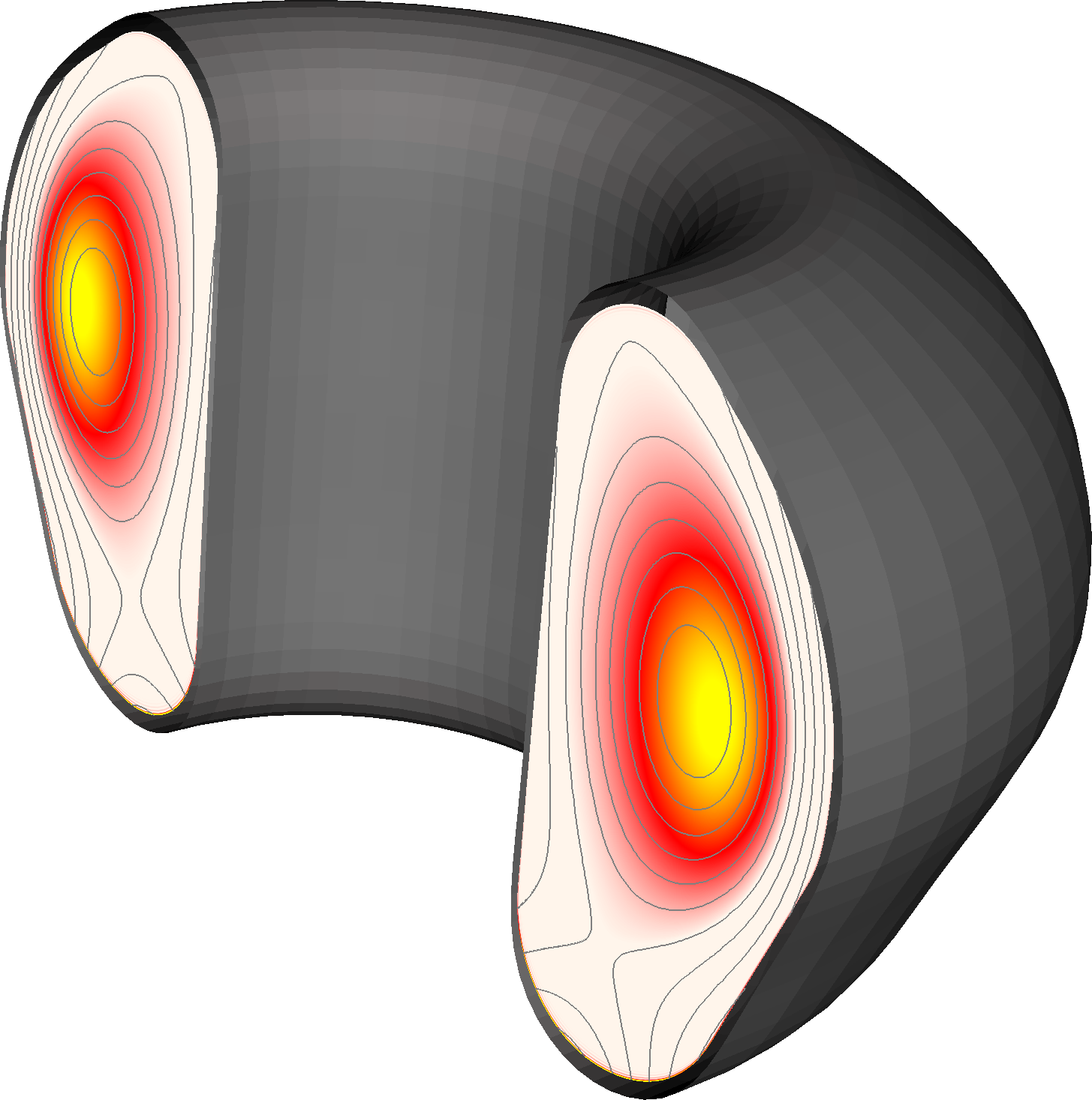}
\hspace{1em}
\includegraphics[width=0.25\textwidth]{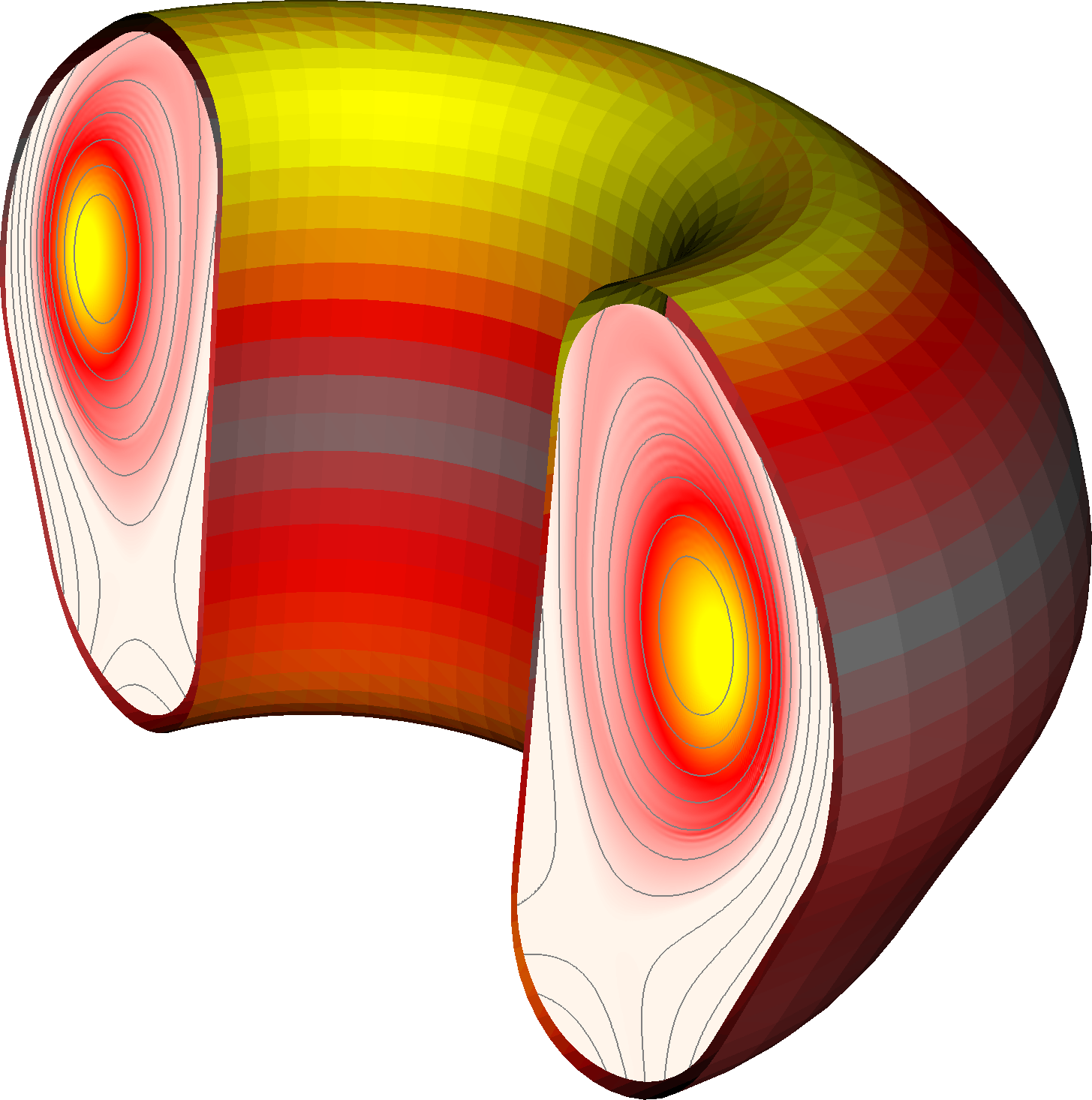}
\hspace{1em}
\includegraphics[width=0.33\textwidth]{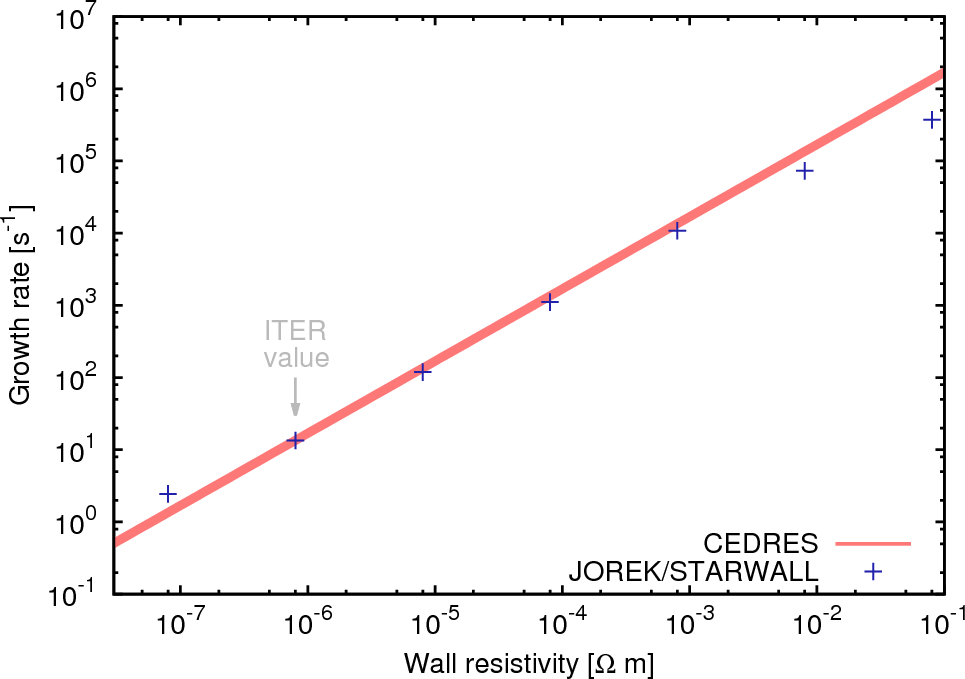}
\caption{ITER-like X-point plasma before (left) and during (middle) a VDE
with an axisymmetric wall. Colors correspond to the plasma current
density (multiplied by $R$ due to normalization) and wall current
density, respectively. The right figure shows the benchmark of linear growth rates
for different wall resistivities against the CEDRES++ code with excellent
agreement. Small deviations are observed at large wall resistivities since CEDRES++
neglects plasma inertia and at extremely small wall resisistivities below the
realistic ITER values due to resolution limitations.
}
\label{fig:vde-bench}
\end{figure}

% ==============================================================================
\section{Benchmarks}\label{:bench}
% ==============================================================================

This section briefly describes benchmarks carried out for the validation of
JOREK-STARWALL. Linear growth rates are compared for resistive wall modes against
analytical theory, and for a vertical displacement event against the CEDRES++ code~\cite{Hertout2011}.
Additional tests have been carried out earlier~\cite{Hoelzl2012A} by reproducing
free-boundary equilibria and comparing tearing mode simulations against a linear
code.

% ==============================================================================
\subsection{Resistive Wall Modes (RWMs)}\label{:bench:rwms}
% ==============================================================================

The growth rates of a resistive wall mode (RWM) are compared~\cite{McAdams2013A,McAdams2014} to analytical
theory~\cite{Liu08A} for a circular equilibrium. As seen in Figure~\ref{fig:rwm-bench}, linear
growth rates show very good agreement except for some deviations at small
plasma-wall distances. Such cases where the resistive wall decreases the growth
rates of the instability to the order of $1\;\mathrm{s}^{-1}$ or even below are hard
to resolve since the natural boundary conditions tend to induce osciallatory
behaviour in these cases. We will identify the reasons and improve this in the
near future. This problem disappears for larger plasma-wall distances or higher
wall resistivities, where growth rates stay higher.

% ==============================================================================
\subsection{Vertical Displacement Events (VDEs)}\label{:bench:vdes}
% ==============================================================================

For an ITER-like VDE testcase with an axisymmetric wall, linear growth rates
obtained with JOREK-STARWALL are compared in Figure~\ref{fig:vde-bench}
to the CEDRES++ code
and show excellent agreement at realistic ITER wall resistivities~\cite{Hoelzl2013}.
For this benchmark, the plasma resistivity needs to be chosen between an upper limit
above which the equilibrium changes too quickly, since we did not use a current
source term, and a lower limit below which eddy currents in the scrape off layer
would become dominant compared to the resistive wall currents.

% ==============================================================================
\section{Current Physics Applications}\label{:phys}
% ==============================================================================

This Section briefly presents first physics applications of JOREK-STARWALL to RWMs,
VDEs, QH-Mode, and disruptions triggered by massive gas injection.

% ==============================================================================
\subsection{Resistive Wall Modes (RWMs)}\label{:phys:rwms}
% ==============================================================================

\begin{figure}
\centering
\includegraphics[width=0.55\textwidth]{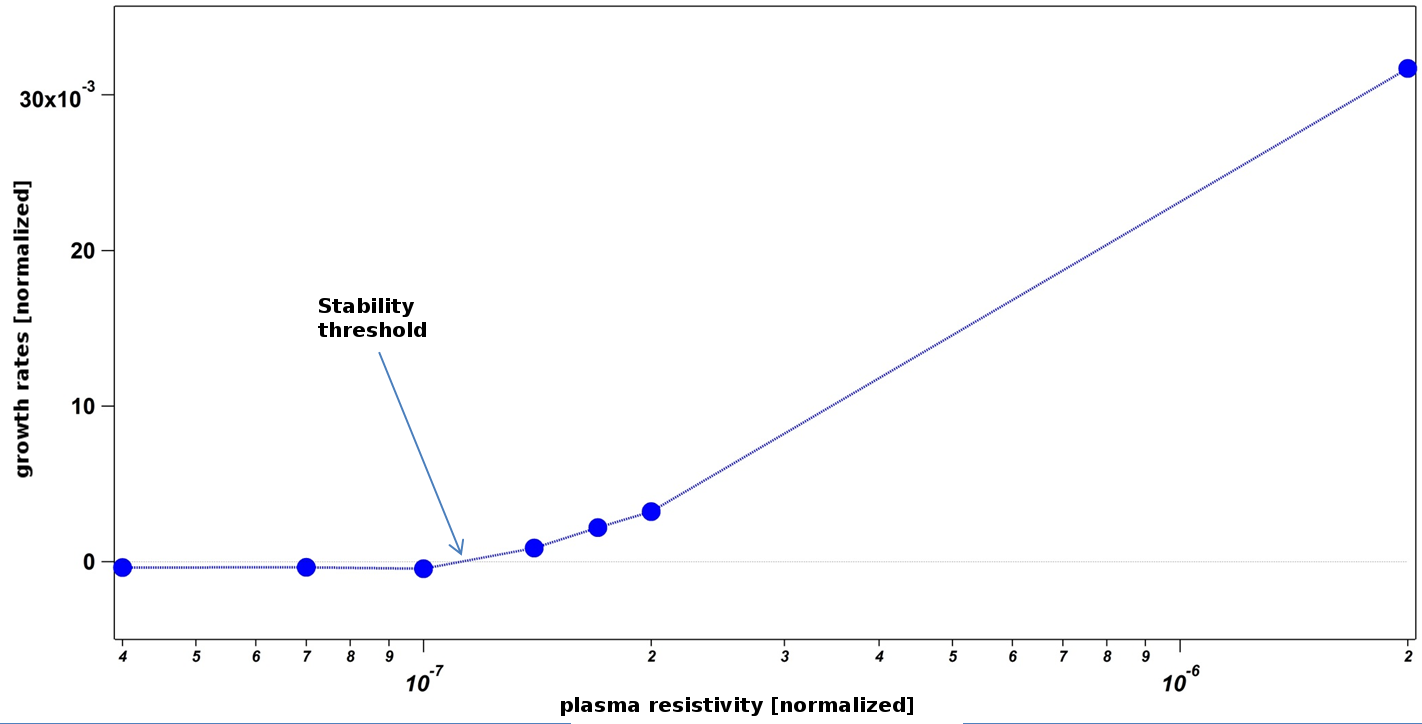}
\caption{Simulation of an RWM in ITER geometry, however at artificially increased
values of resistivity and viscosity due to computational limitations. The growth
rate of the observed instability shows a strong dependence on the plasma resistivity
indicating a resistive instability. It is fully stabilized below a threshold in the
plasma resistivity. Below this value, stabilizing dissipative effects (like
viscosity) become dominant over the growth rate of the resistive instability.
}
\label{fig:iter-rwm}
\end{figure}

Resistive wall modes were simulated in an ITER scenario with reversed
q-profile~\cite{McAdams2014}. Results with the realistic location of the conducting wall
in ITER were compared to an artificial case with closer wall, and the no-wall case.
The comparison of those cases, and additional wall resistivity scans allows us to quantify
the stabilizing effect of the walls. Figure~\ref{fig:iter-rwm} shows a
threshold observed in the plasma resistivity, below which the mode is fully
stabilized. Consequently, this mode is a resistive mode. 
The existence of the threshold and the resistivity value it is observed at
depend on the considered equilibrium, wall configuration, and plasma parameters.

% ==============================================================================
\subsection{Vertical Displacement Events (VDEs)}\label{:phys:vdes}
% ==============================================================================

%data: /ptmp1/work/mhoelzl/data/2014-07-24_ksenia_vde_3d
\begin{figure}
\centering
\includegraphics[width=0.24\textwidth]{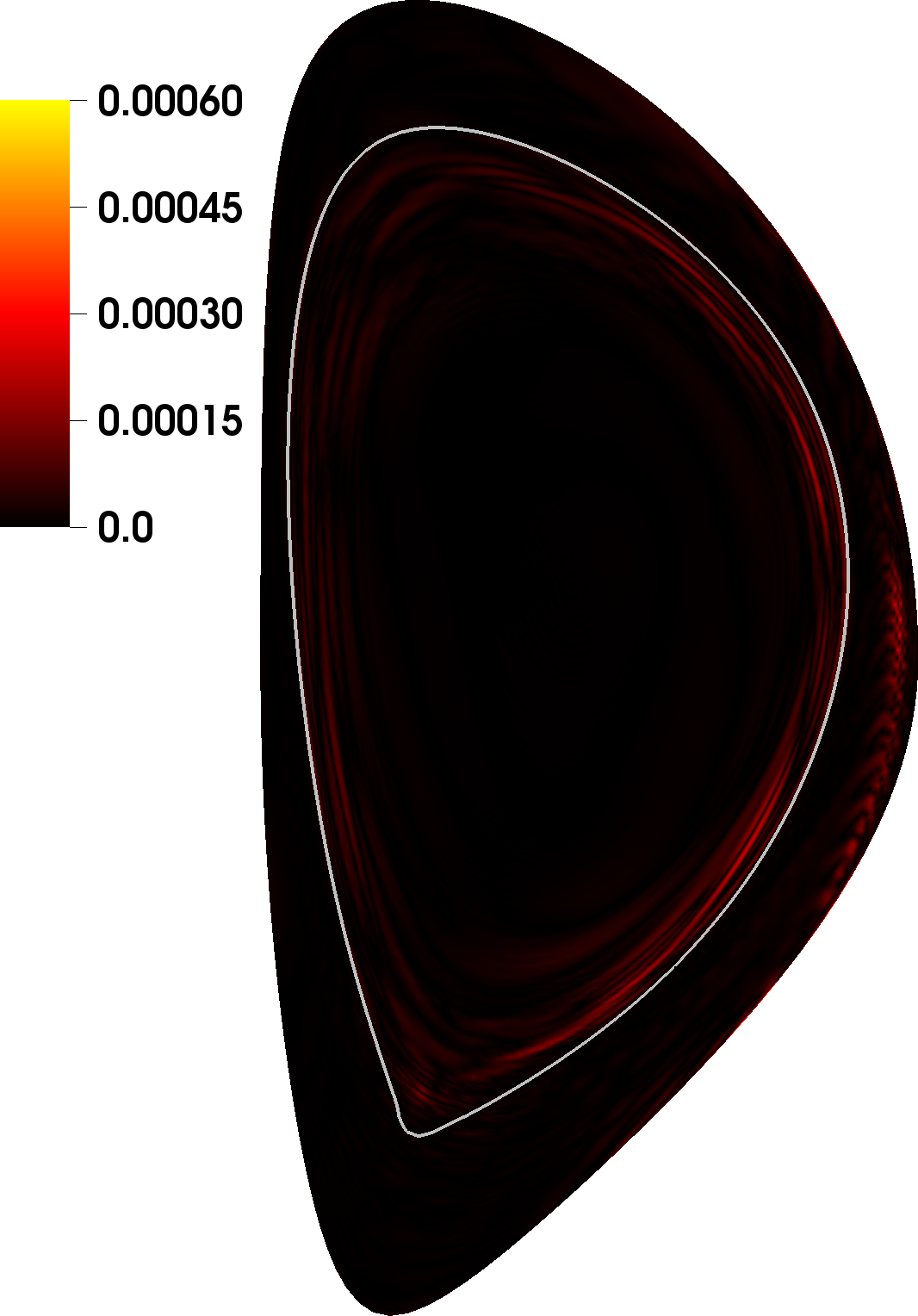}
\hspace{1em}
\includegraphics[width=0.24\textwidth]{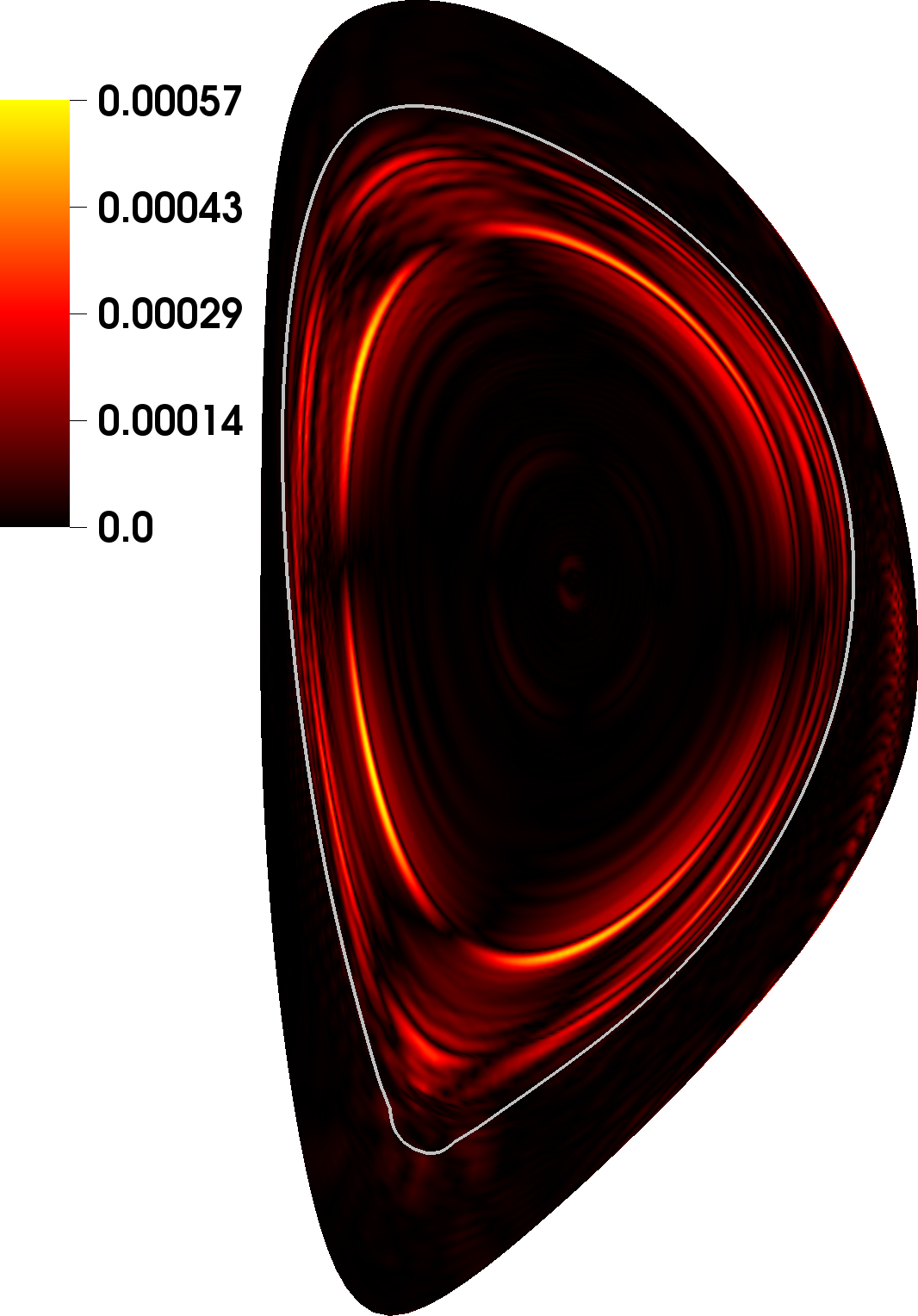}
\hspace{1em}
\includegraphics[width=0.24\textwidth]{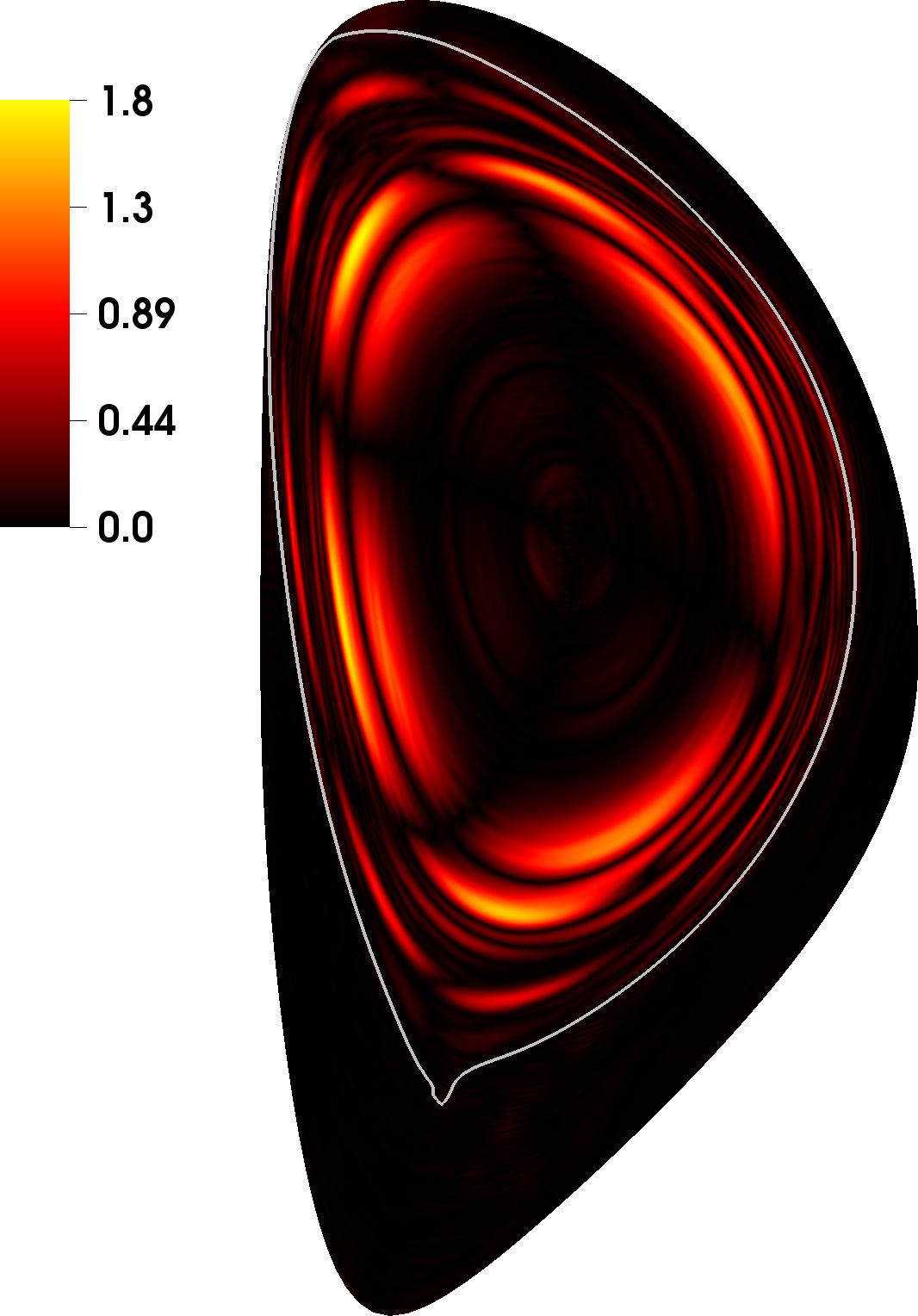}
\caption{The absolute value of the $n=1$ component of the current density is plotted (in
normalized units) close to the beginning
of the simulation where the instability is almost axisymmetric (left, $t=40\;t_A$), around the onset
of the $n=1$ perturbation where the axis has already moved upwards by about $1\;\mathrm{cm}$ due to the
vertical instability (middle, $t=1200\;t_A$) and at a later stage where the vertical axis displacement is about $32\;\mathrm{cm}$
(right, $t=9600\;t_A$). The grey contour line corresponds to the pedestal temperature indicating the location of the
separatrix. Clearly, the plasma has changed its shape in the course of the instability and the evolving $n=1$ mode
has become much less radially localized.}
\label{fig:iter-vde}
\end{figure}

A vertical displacement event is an axisymmetric instability where the whole plasma
moves up- or downwards with an approximately exponential evolution of the displacement in time. The time
scale is governed by geometry and conductivity of conducting structures. A non-axisymmetric
component can arise during the non-linear evolution of the VDE (external kink) when the plasma
begins to shrink and the q-profile changes. Figure~\ref{fig:iter-vde} shows
a simulation of a VDE in ITER geometry (however with increased wall and plasma resistivities)
which develops such a non-axisymmetric feature~\cite{Ksenia2014}. The vertical
upward movement of the initially axisymmetric plasma is triggered by a small
perturbation of coil currents. A small $n=1$ perturbation at
the separatrix can be observed (left Figure). A bit later, a $2/1$ mode sets in at the $q=2$
surface while some perturbations between this surface and the separatrix are still visible
(middle). The $2/1$ mode continues to grow and gets radially less localized while the
vertical movement of the plasma continues (right).

% ==============================================================================
\subsection{Quiescent H-Mode (QH-Mode)}\label{:phys:qh}
% ==============================================================================

\begin{figure}
\centering
\includegraphics[height=0.23\textwidth]{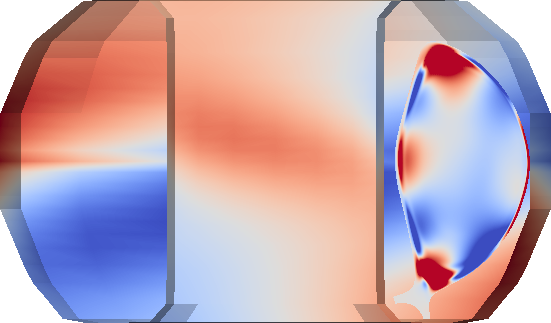}
\hspace{1em}
\includegraphics[height=0.23\textwidth]{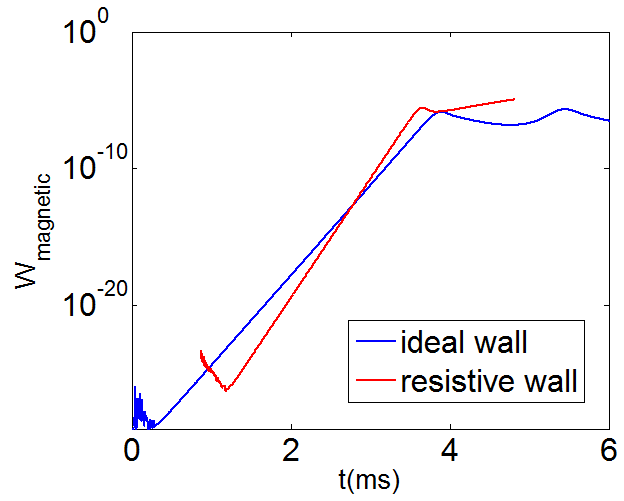}
\caption{The perturbed flux and wall potential on the resistive wall are plotted
for a saturated kink/peeling mode for a simulation of DIII-D shot \#153440 (left) with
JOREK-STARWALL. The time evolution
of the $n=1$ magnetic energy (right) shows that the mode goes into a
saturated state. Both, the initial growth rate as well as the saturation level are
higher with the resistive wall boundary conditions of JOREK-STARWALL compared to ideal wall
boundary conditions at the boundary of the JOREK computational domain.}
\label{fig:qh}
\end{figure}

Quiescent H-Mode has been obtained in DIII-D~\cite{Burrell2002}, ASDEX Upgrade~\cite{Suttrop2005}
and other devices. It is characterized by a saturated kink/peeling mode~\cite{Snyder2007}
and does not feature periodic ELM crashes.
The influence of the vacuum vessel wall on the destabilization and saturation of
edge modes for DIII-D QH-mode plasma has been studied with JOREK-STARWALL.
Figure~\ref{fig:qh} shows the flux perturbation and the wall current stream function
in the saturated state.
When comparing to a fixed boundary simulation (i.e., ideal wall
at the JOREK boundary; see Fig.~\ref{fig:qh} left) the saturated amplitude of the kink/peeling mode is larger
in the ``free boundary'' simulation (resistive wall outside the JOREK domain; see Fig.~\ref{fig:qh} left).
The initially rotating mode gets locked to the wall in the
saturated phase of those simulations. To reproduce experimental observations in more
detail, toroidal rotation needs to be taken into account (partly done already in
Reference~\cite{Liu2014b}), and also effects from neoclassical poloidal and diamagnetic
rotation are important (simulations in preparation).

% ==============================================================================
\subsection{Disruption Events}\label{:phys:disr}
% ==============================================================================

\begin{figure}
\centering
\includegraphics[width=0.34\textwidth]{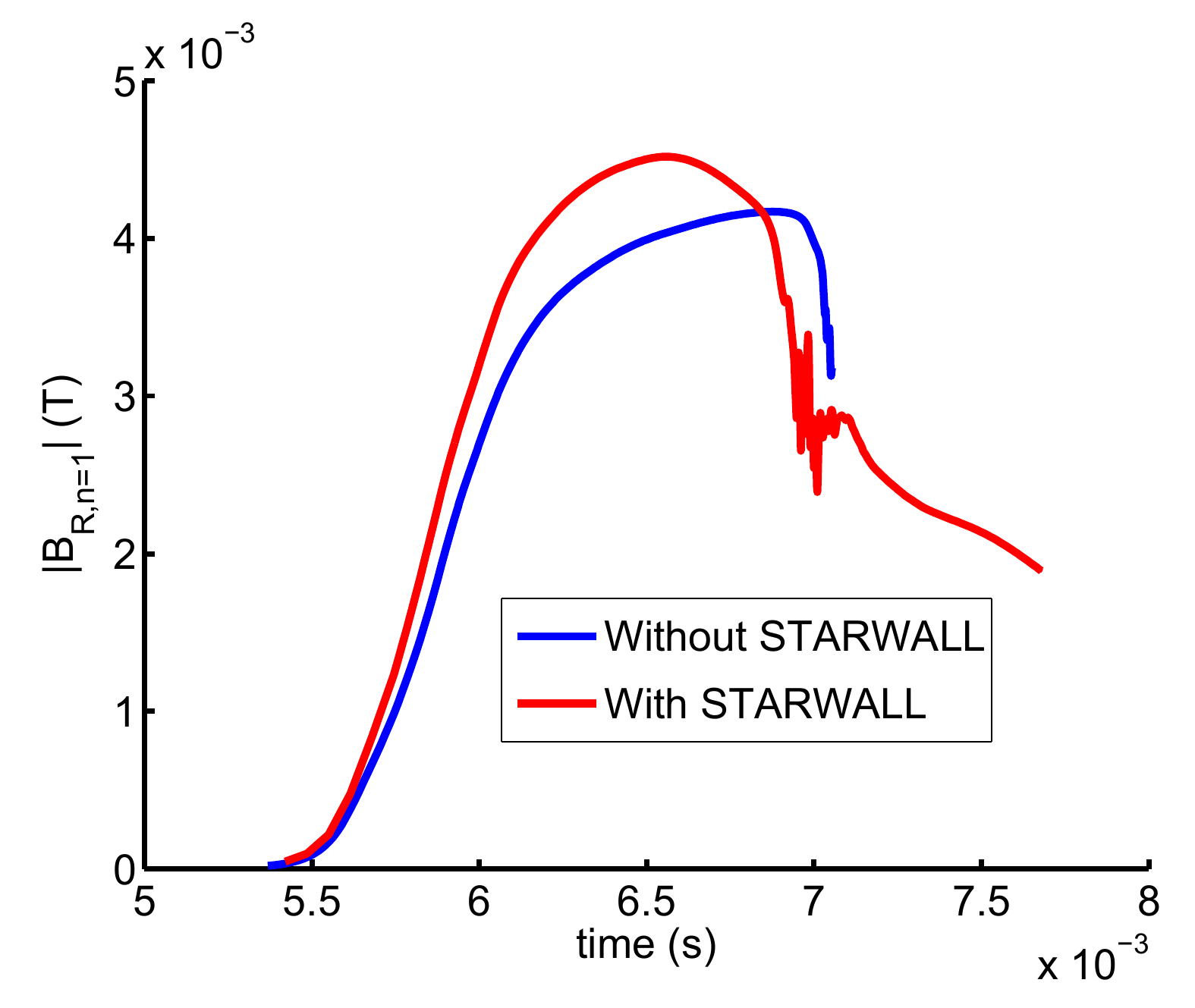}
\hspace{1em}
\includegraphics[width=0.34\textwidth]{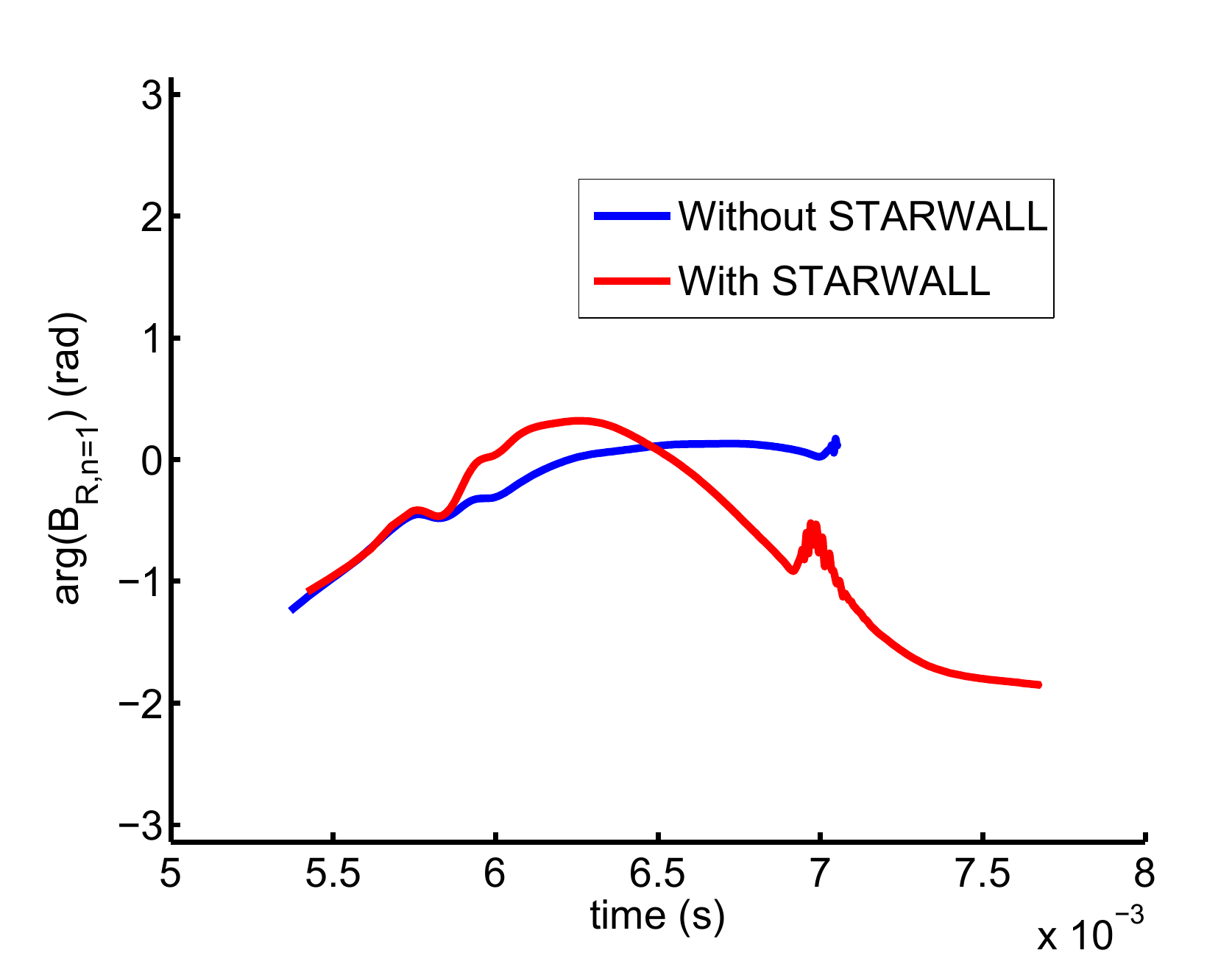}
\caption{Amplitude (left) and phase (right) of the synthetic locked mode signal
produced as a post-processing of JOREK simulations of MGI in JET. In our simulations,
the massive gas injection (currently assuming a neutral source at zero momentum) probably is the
main cause for the locking of the MHD activity, since the electro-magnetic force
responsible for mode locking vanishes in the ideal wall case.}
\label{fig:mgi}
\end{figure}

JOREK simulations of disruption mitigation by Massive Gas Injection (MGI) are in
progress~\cite{Fil2014}. The focus at the moment is on the cooling phase (when the gas enters the
plasma and generates a cold front progressing from the edge to the core) and the
triggering and development of the thermal quench (TQ). The MGI is modeled as a poloidally
localized (and due to the low toroidal resolution of our first simulations toroidally weakly localized)
source of neutrals which are assumed to move only by diffusive processes. Simulations with JOREK standalone
(without STARWALL) for a JET plasma showed that an m=2, n=1 mode is triggered when the
cold front reaches the q=2 surface, and that the TQ ensues when an m=3, n=2 mode is
destabilized~\cite{Fil2014}. These simulations have been repeated with STARWALL (for the
n=1 harmonic only). Even though the behavior is qualitatively similar, these simulations
show quantitative differences and will allow to study, among other things, the dynamics of
the locked mode (LM). Indeed, in JET disruptions, a LM is systematically observed. The LM
signal is one of the main elements that can be exploited for disruption prediction~\cite{DeVries2014},
hence it is important to understand LM physics and its relation with
disruptions. Figure~\ref{fig:mgi} shows synthetic LM signals constructed by post-processing the
above-mentioned JOREK simulations for JET with and without STARWALL. The left (resp.\ right)
figure shows the amplitude (resp.\ phase) of the LM signal, which is defined as the
n=1 component of the radial magnetic field at the machine midplane. The TQ sets in at about $7\;\textrm{ms}$,
when the signals appear to be noisy. One can see that the LM amplitude peaks slightly
before the TQ, with a similar peak value with and without STARWALL, on the order of $4\;\textrm{mT}$.
Experimentally, the LM amplitude also typically presents a peak, although usually it is
slightly after the TQ and its magnitude is smaller, typically $0.5\dots2\;\textrm{mT}$ [De Vries EPS 2014].
The LM phase shows a different behavior with and without STARWALL, probably due to the
modified torque exerted by the resistive wall on the plasma. These first MGI simulations do not
pretend to be quantitatively correct, for several reasons: only $n=0\dots2$ harmonics are
treated, the atomic physics is simplified, the neutral source is constant in time instead
of having a realistic shape, etc. Present work is dedicated to improving all these points
in order to simulate as realistically as possible a discharge with a D$_2$ MGI in JET, and to
compare the results with experimental measurements in order to validate the model.
JOREK-STARWALL will also be essential to study current quench and VDE phases. 

% ==============================================================================
\section{Perspectives for an Extension to Halo Currents}\label{:halo}
% ==============================================================================

Halo currents arising during a disruption event are a major concern for ITER due
to the strong rotating asymmetric forces acting on vessel structures they may cause.
Non-linear modeling will
be important to predict these forces and also the effectiveness of disruption mitigation
techniques aiming to reduce them. The JOREK-STARWALL code, however, currently assumes
the wall surface currents to be divergence free, which allows us to include only eddy
current effects.
This Section provides a brief outlook to a future extension of JOREK-STARWALL to halo
currents which effectively are current sources or sinks for the wall currents.

In the present STARWALL code, wall currents are given by $\mathbf{j}_\text{w} = \hat{n}\times\nabla\phi_\text{w}$
where the current stream functions $\phi_\text{w}$ are discretized by values at each triangle
node and $\hat{n}$ denotes the normal vector at each triangle. In order to describe
halo currents, which effectively enter as sources and sinks for the wall currents,
this expression has to be generalized~\cite{Atanasiu2013} to
\begin{equation*}
\mathbf{j}_\text{w} = \hat{n}\times\nabla\phi_\text{w} + \nabla\varphi_\text{w},
\end{equation*}
by adding a second non-divergence free surface current component, where $\varphi_\text{w}$
takes the role of the electric potential on the wall.

In the derivation of the STARWALL equations, the starting point is a Lagrangian describing
the energy associated to the vacuum fields and wall currents which involves terms depending
on virtual surface currents (representing the plasma side) and the poloidal magnetic flux
at the boundary of the JOREK domain as well as terms depending on the wall currents.
A set of time evolution equations can be obtained by applying a variational principle and
eliminating the virtual surface currents from the system. The coefficients of the
equations can be written in terms of matrices (see Ref.~\cite{Hoelzl2012A}) and are used in
JOREK to evolve the wall currents in time.
The generalization of the wall current
expression will extend the system of equations, i.e., introduce additional
matrices and increase the size of some existing matrices. To make further required
extensions clear, let us briefly summarize some technical details of the JOREK-STARWALL
coupling:
\begin{enumerate}
\item The JOREK computational grid is constructed and the information describing the
boundary of the computational domain is written out, i.e., the boundary nodes and elements.
\item STARWALL calculates the "vacuum + wall response" matrices based on this JOREK boundary information
and the desired wall geometry. Another JOREK simulation with a different computational
grid requires a new STARWALL run.
\item The JOREK time evolution is carried out in free boundary mode using the STARWALL
response matrices to evolve wall currents and to calculate the
tangential magnetic field at the boundary of the JOREK domain. The tangential field
is required for a boundary integral in the current definition equation arising from
partial integration which acts as a natural boundary condition~\cite{Hoelzl2012A}.
\end{enumerate}

When introducing halo currents, consistency needs to
be guaranteed between plasma and wall for currents flowing between both (current conservation).
Additionally, the electric potential must be consistent between
plasma and walls. In practice, STARWALL will provide a matrix correlation between
currents flowing into the wall and electric potential. This correlation
will be used as a boundary condition in JOREK. Although the reduced MHD model of JOREK
only has a variable for the toroidal current density, the full current density can be
reconstructed (\ref{:app:red}) such that halo currents will be compatible with, both, reduced
and full MHD models.

A different longer term option would be to implement 3D conducting walls with finite thickness
directly into the JOREK domain. A planned future implementation of toroidal finite elements
in JOREK (instead of the present toroidal Fourier expansion) will clear the way for this option.
STARWALL would then be used for free boundary conditions and conducting structures further
away from the plasma. This approach will be tested soon by first implementing an axisymmetric
conducting wall inside the JOREK domain.

% ==============================================================================
\section{Summary}\label{:concl}
% ==============================================================================

We have presented some benchmarks of the coupled non-linear MHD code JOREK and
the resistive wall code STARWALL against linear codes and analytical theory. Some
first physics applications to RWMs, VDEs, QH-mode and disruptions were also shown.

An outlook is given to a future extension of JOREK-STARWALL to include halo currents.
This will, together with separate other ongoing physics development projects in JOREK, allow
for non-linear simulations of the most important processes relevant for disruption
physics in the future.

% ==============================================================================
\section*{Acknowledgements}
% ==============================================================================

This work was carried out using the HELIOS supercomputer system at Computational
Simulation Centre of International Fusion Energy Research Centre (IFERC-CSC),
Aomori, Japan, under the Broader Approach collaboration between Euratom and Japan,
implemented by Fusion for Energy and JAEA.

% ==============================================================================
\section*{References}
% ==============================================================================

\bibliography{mybib}

% ==============================================================================
% ==============================================================================
\appendix
% ==============================================================================
% ==============================================================================

% ==============================================================================
\section{Validity of the thin wall approximation}\label{:app:thin}
% ==============================================================================

STARWALL uses the thin wall approximation for 3D resistive wall
structures. Conductors are discretized by triangles. The
resistivity acting on the surface currents under this assumption is
$\eta_\text{surf}=\eta/d$ where $\eta$ is the specific resistivity of the wall
material. We will discuss the validity of the thin wall approximation
in the following, showing that it is a reasonable model in most
situations. For the discussion, we will consider a typical time scale
$\tau$ of the perturbation, which can either be the inverse of
its rotation frequency or the inverse of an instability growth rate.

A magnetic field varying on the characteristic time scale $\tau$
penetrates into a conductor limited to the skin depth $\delta_\text{skin} =
\sqrt{ 2 \eta \tau / (2 \pi \mu_r \mu_0)}$, where $\mu_0$ denotes the vacuum
permeability, and $\mu_r$ the relative permeability. ITER wall structures are
mainly built from austenitic steel, and even martensitic components
will be deep in the saturated regime due to the large toroidal field, such that
$\mu_r=1$ can be taken in estimates. For the steel walls ITER will have ($\eta \approx
10^{-6}\;\Omega\;\mathrm{m}$) we obtain
  $\delta_\text{skin} \approx 0.5\mathrm{m}\cdot\sqrt{\tau[\mathrm{s}]}$.
The characteristic time for a magnetic perturbation to become homogeneous across
the $d=6 \mathrm{cm}$ thick ITER walls is given by
  $\tau_\text{skin} = (0.06 \mathrm{m})^2 / (0.5 \mathrm{m}^2/\mathrm{s}) = 7 \mathrm{ms}$.

An external kink mode which can fully be stabilized by an ideally
conducting wall, but is unstable in the presence of a resistive wall
is called a resistive wall mode. Its growth time $\tau$ is approximately
given by the characteristic $L/R$ time $\tau_\text{wall}$ of the wall structures,
which for the ITER conducting vessel is of the order of $200 \mathrm{ms}$. For
such a mode with $\tau \approx \tau_\text{wall} \gg \tau_\text{skin}$, the field is homogeneous
across the wall and the thin wall approximation is well justified.

The opposite limit of an instability that is weakly coupled to the
walls and can not be stabilized by ideally conducting walls
corresponds to $\tau \ll \tau_\text{skin} \ll \tau_\text{wall}$. In this case, the magnetic
perturbation can only penetrate into a very thin surface layer of the
conductors. In principle, an effective -- increased -- resistivity needs to
be considered for the wall current evolution:
$\eta_\text{surf,eff} = \eta/\delta_\text{skin}$ instead of $\eta_\text{surf} = \eta/d$.
However, the wall completely shields the magnetic field associated to
such a fast mode and effectively acts as an ideally conducting wall
-- however with little influence on the mode growth.
So a correction for the resistivity is not necessary to
describe the dynamics of the mode, the thin wall
approximation is well justified.

The remaining case would be an instability with $\tau \approx \tau_\text{skin}$ where
the magnetic perturbation associated to the mode has partially
penetrated through the wall. This situation corresponds to an increased wall
resistivity $\eta_\text{surf,eff}$, which now really has to be taken into
account in order to keep the thin wall approximation valid. The non-linear behaviour
of an instability might not be obtained fully accurately with the thin-wall
approximation in this case. Tokamak instabilities with such growth
rates are, however, rather untypical since the wall coupling must be very
marginal.

The geometrical simplification by the thin wall only plays
a role if the mode wall distance is smaller than the wall thickness. In such an unusual case
(e.g., large passive stabilizing loop present in ASDEX Upgrade), conductors
can be discretized by several layers of thin walls.

% ==============================================================================
\section{Halo currents with reduced MHD}\label{:app:red}
% ==============================================================================

In the JOREK reduced MHD model, only a toroidal current component exists as a
variable. Since the boundary of the JOREK computational domain is axi-symmetric,
no currents could enter the walls preventing halo current modeling.
However, the full current vector can be reconstructed assuming low pressure at
the JOREK boundary: The momentum conservation equation then turns into
$\mathbf{j}\times\mathbf{B}=0$, i.e., the plasma current is parallel to the
magnetic field. The current component flowing into the wall is then given by
$j_\bot=j\mathbf{b}\cdot\mathbf{\hat{n}}$, where $j$ denotes the toroidal current
component in JOREK, $\mathbf{b}=\mathbf{B}/B$, and $\mathbf{\hat{n}}$ is the
unit vector perpendicular to the conducting wall.

\end{document}